# A study of global monopole in Lyra geometry

F. Rahaman, S.Mal and P. Ghosh

Department of Mathematics
Jadavpur University, Kolkata – 700 032, India

E-mail: farook_rahaman@yahoo.com

**Abstract:**

A class of exact static solution around a global monopole resulting from the breaking of a global S0(3) symmetry is obtained in the context of Lyra geometry.
Our solution is shown to possess an interesting feature like 'wormholes' space-time. It has been shown that the global monopole exerts no gravitational force on surrounding non-relativistic matter.

The origin of structure in the universe is one of the greatest cosmological mysteries even today. Extended topological objects such as monopoles, strings and domain walls may play a fundamental role in the formation of our universe [1]. Phase transitions in the early universe can give rise to these topological defects. A topological defect is a discontinuity in the vacuum and can be classified according to the topology of the vacuum manifold. Monopoles are point like topological defects and are formed where M contains surfaces which cannot be continuously shrunk to a pointy i.e. when $\pi_2 (M) \neq I$. ( M is the vacuum manifold ) [2]. Global monopoles are formed from breaking of a global S0(3) symmetry. Such monopoles have Goldstone fields with energy density decreasing with the distance as inverse square law. They are also found to have some interesting features in the sense that a monopole exerts no gravitational force on its surrounding non-relativistic matter but the space-time around it has a deficit solid angle.

At first, Barriola and Vilenkin (BV) [3] showed the existence of such a monopole solution resulting from the breaking of global S0(3) symmetry of a triplet scalar field in a Schwarzchild back ground. After that so many works have been done on monopole [4].
The above discussions have been confined within the context of General Relativity.
In last few decades there has been considerable interest in alternative theories of gravitation. The most important among them being scalar-tensor theories proposed by Lyra [5] and Brans-Dicke [5]. Lyra suggested a modification of Riemannian geometry, which may also be considered as a modification of Weyl's geometry.

In Lyra's geometry, Weyl's concept of gauge, which is essentially a metrical concept, is modified by the introduction of a gauge function into the structure-less manifold.

In general relativity Einstein succeeded in geometrising gravitation by identifying the metric tensor with gravitational potentials.

In the scalar tensor theory of Brans-Dicke on the other hand, the scalar field remains alien to the geometry. Lyra's geometry is more in keeping with the spirit of Einstein's principle of geometrisation since both scalar and tensor fields have more or less intrinsic geometrical significance.

In consecutive investigations Sen [6] and Sen and Dunn [7] proposed a new scalar tensor theory of gravitation and constructed an analogue of the Einstein field equations based on Lyra's geometry which in normal gauge may be written as

$$R_{ab} - \tfrac{1}{2} g_{ab} R + (3/2) \varphi_a \varphi_b - \tfrac{3}{4} g_{ab} \varphi_c \varphi^c = -8\pi T_{ab} \quad \ldots(1)$$

where $\varphi_a$ is the displacement vector and other symbols have their usual meaning as in Riemannian geometry.

Subsequent investigations were done by several authors in scalar tensor theory and cosmology within the framework of Lyra geometry [8].

In recent, we have studied some topological defects within the framework of Lyra's geometry [9].

In this paper, we would like to discuss exact monopole solutions based on Lyra's geometry in normal gauge i.e. displacement vector

$$\varphi_i = (\beta(r), 0, 0, 0) \quad \ldots\ldots(2)$$

and look forward whether the monopole shows any significant properties due to introduction of the gauge field in the Riemannian geometry.

Since the space-time of global monopole is static and spherically symmetric, the metric is taken as

$$ds^2 = e^\gamma dt^2 - e^\mu dr^2 - r^2 d\theta^2 - r^2 \sin^2\theta d\varphi^2 \quad \ldots\ldots(3)$$

where $\gamma, \mu$ are functions of r alone. In the original work of Barriola and Vilenkin, the energy momentum tensor due to the monopole field outside the core be [3]

$$T^t_t = T^r_r = (\eta^2/r^2); \quad T^\theta_\theta = T^\varphi_\varphi = 0 \quad \ldots(4)$$

where $\eta$ is the symmetry breaking scale of the theory.

With the equations (3) and (4), the equation (1) gives explicitly the following relations.

$$e^{-\mu} [(\gamma^1 / r) + (1/r^2)] - (1/r^2) + \tfrac{3}{4} e^{-\gamma} \beta^2 = -(8\pi\eta^2/r^2) \qquad ..(5)$$

$$-e^{-\mu} [(\mu^1/r) - (1/r^2)] - (1/r^2) - \tfrac{3}{4}\beta^2 e^{-\gamma} = -(8\pi\eta^2/r^2) \qquad ....(6)$$

$$e^{-\mu} [\tfrac{1}{2} \gamma^{11} + \tfrac{1}{4} (\gamma^1)^2 - \tfrac{1}{4} \mu^1\gamma^1 + \tfrac{1}{2} \{(\gamma^1 - \mu^1)/r\}] + \tfrac{3}{4} \beta^2 e^{-\gamma} = 0 \qquad ....(7)$$

('$^1$' indicates the differentiation w. r. t. 'r').

By using the following combination (5) – (6) – 2.(7), one can get

$$\gamma^{11} + 2(\gamma^1/r) + \tfrac{1}{2} (\gamma^1)^2 = \tfrac{1}{2} \gamma^1\mu^1 \qquad .......(8)$$

One may attempt to eq.(8) for two different situations:

(i) $\gamma^1 = 0$, which means $\gamma$ = constant.

(ii) $\gamma^1 \neq 0$, which in turn leads to the differential equation,

$$e^{-\gamma} (\gamma^1)^2 = b^2 e^{\mu} r^{-4} \qquad ......(9)$$

where b is an arbitrary integrating constant. In the first case, one may choose $\gamma = 0$, without any loss of generality. In the second case $\gamma^1 \neq 0$, it is extremely hard task to obtain the solutions to the field equations. Hence we abandon this case in our present work and proceed with the other case i.e. $\gamma = 0$ for further calculations. With $\gamma = 0$, one gets, after some simple and straight forward calculations the following two equations:

$$e^{-\mu} [(2/r^2) - (\mu^1/r)] = (2/r^2) - (16\pi\eta^2/r^2) \qquad ..(10)$$

$$-e^{-\mu} [(\mu^1/r)] = (3/2)\beta^2 \qquad ......(11)$$

Solving the above two equations, we get

$$e^{-\mu} = 1 - 8\pi G\eta^2 - [D/r^2] \qquad ..(12)$$

$$\beta^2 = (4D/3) r^{-4} \qquad ....(13)$$

where D is an integration constant.

Hence the monopole solutions read

$$ds^2 = dt^2 - [1 - 8\pi G\eta^2 - D r^{-2}]^{-1} dr^2 - r^2( d\theta^2 + \sin^2\theta \, d\varphi^2) \qquad ...(14)$$

[ Here $1 - 8\pi G\eta^2 > 0$. Since for a typical grand unified theory the parameter η is of order $10^{16}$ GeV. So $8\pi G\eta^2 \sim 10^{-5}$ ].

Now depending on the sign of D, we have the following two cases.

**Case – I: D > 0**

The equation of geodesics for the monopole metric given by (14)

$$r^{\cdot 2} \equiv (dr/ds)^2 = [1 - 8\pi G\eta^2 - Dr^{-2}][E^2 - (J^2/r^2) - L] \qquad \ldots(15)$$

$$\varphi^{\cdot} \equiv (d\varphi/ds) = (J/r^2) \qquad \ldots(16)$$

$$t^{\cdot} \equiv (dt/ds) = E \qquad \ldots(17)$$

where the notion is as usual considered to be taking place in the $\theta = (\pi/2)$ planes and the constants E & J having the respective interpretations of energy per unit mass and angular momentum about an axis perpendicular to the invariant plane $\theta = (\pi/2)$. Here s is an affine parameter and L is the Lagrangian having values 0 and 1 respectively for null and time like particles.

Now the equation for radial geodesics ( J = 0 ):

$$(dr/dt)^2 = [1 - 8\pi G\eta^2 - Dr^{-2}][1 - LE^{-2}] \qquad \ldots(18)$$

with solution

$$t = \pm [\{(1 - 8\pi G\eta^2)r^2 - D\}/\{(1 - LE^{-2})\}]^{\frac{1}{2}} + \text{constant} \qquad \ldots(19)$$

and the affine parameter $s \propto t$.

The eq.(19) represents a hyperbola and shows that to an external observer a radially in-falling time like or null particle approaches the radius $r = [D/(1 - 8\pi G\eta^2)]^{\frac{1}{2}}$ asymptotically but can never reach it. Here we also see the s – r relationship represents a hyperbola. Now, for time like geodesics ( L = 1 ), s is the proper time and hence an observer falling with time like particle also skirts the physical singularity at r = 0 by asymptotically grazing the critical radius at
$r = [D/(1 - 8\pi G\eta^2)]^{\frac{1}{2}}$. This feature is characteristic of a wormhole space-time [10].

**Case – II: D < 0.**

For D < 0, the solution is well behaved everywhere except at r = 0 where it encounters the physical singularity. This case is not realistic since β takes imaginary value for D < 0.

Before ending our study we would like to point out the gravitational force acts on surrounding the monopole. The radial component of the acceleration acting on a test particle in the gravitational field of the monopole is given by

$$A^r = V^1{}_{;0} V^0.$$

For a co moving particle $V^a = [1/\sqrt{(g_{00})}]\delta_0{}^a$.

Hence using the line element (14) one can calculate $A^r$ which becomes

$$A^r = 0 \qquad \qquad \ldots(20)$$

Hence the monopole exerts no gravitational influence on the matter around it.

In conclusion, this work extends the earlier work of BV to scalar tensor theory based on Lyra geometry. We have found an exact of solution for the metric outside the monopole core resulting from the breaking of a global S0(3) symmetry. An interesting feature of this solution is that under a certain condition, the solution corresponds to ' wormhole ' in space-time with throat radius
$r = [D/(1 - 8\pi G\eta^2)]^{1/2}$.
This result implies that space-time of global monopole within the frame work of Lyra geometry induces a wormhole geometry in space-time.
Our monopole (exact solution) exerts no gravitational force on non-relativistic particles. This result is in agreement with BV monopole.

## Acknowledgements:

One of the authors (FR) is thankful to IUCAA for providing the research facility.
We are grateful to the referee for pointing out the errors in the earlier version.